\documentstyle[epsfig,preprint,aps]{revtex}  
\input{epsf} 
 
\newcommand{\vub}{V_{\rm ub}} 
\newcommand{\vcb}{V_{\rm cb}} 
\newcommand{\vcs}{V_{\rm cs}}
\newcommand{\be}{\begin{eqnarray}} 
\newcommand{\ee}{\end{eqnarray}} 

\newcommand{\g}{\gamma}

\renewcommand{\to}{\rightarrow} 
\def\BBbar{B\overline{B}}

\begin{document} 
 
\draft 
 
\setcounter{page}{0} 
 
\begin{titlepage} 
 \begin{flushright} 
 \hfill{\bf (hep-ph/0010157)}\\ 
 \hfill{YUMS 00-08}\\ 
 
 \end{flushright} 
 
\begin{center} 
{\large\bf Measurement of $|V_{ub}/V_{cb}|$ (and $|V_{ub}|$)  
in Exclusive Nonleptonic Decays, $\overline{B}^0\to D_s^{(*)-}(\pi^+,\rho^+)$ and 
$\overline{B}^0\to D_s^{(*)-}D^{(*)+}$ 
} 
\end{center} 
\vskip 0.4cm 
 
\begin{center} 
{\sc   C.~ S.~ Kim$^{~\mathrm{a,b}}$, 
 Y.~ Kwon$^{~\mathrm{a}}$, 
 Jake~ Lee$^{~\mathrm{a}}$ and 
 W.~ Namgung$^{~\mathrm{c}}$   } 
 
 
\begin{small}  
$^{\mathrm{a}}$ {\it Department of Physics and IPAP, Yonsei 
University, Seoul 120-749,  Korea} \\ 
 
$^{\mathrm{b}}$ {\it Department of Physics, University of Wisconsin, 
Madison, WI 53706, USA} 
 
$^{\mathrm{c}}$ {\it Department of Physics, Dongguk University, Seoul 100-715, 
Korea} 
 
\end{small} 
\end{center} 
 
\begin{center} 
 (\today) 
\end{center} 

\vspace{-0.5cm} 

\setcounter{footnote}{0} 
\begin{abstract} 
We have studied extracting $|\vub/\vcb|$ by calculating the ratios  
${\cal B}(\overline{B}^0\to D_s^{(*)-}(\pi^+,\rho^+))/{\cal  
B}(\overline{B}^0\to D_s^{(*)-}D^{(*)+})$ including penguin effects within  
the factorization assumption.  The ratios involving $\overline{B}^0\to  
D_s^-D^{+}$ mode have considerable penguin corrections ($\sim 15\%$ at  
the amplitude level), but those involving $\overline{B}^0\to  
D_s^-D^{*+}$ mode have relatively small penguin corrections.  On the  
other hand, the $\overline{B}^0\to D_s^-D^{+}$ mode has smaller  
form-factor dependance.  Therefore, these ratios complement each other  
in measuring $\vub/\vcb$.  The theoretical uncertainty from the hadronic form 
factors in our method is at the  level of 15\%, which is comparable 
to the model-dependence  uncertainty of about 20\% in the measurement of 
$\left| \vub/\vcb  \right|$ from the exclusive semileptonic $B$ decays.  Using 
the newest upper limit  on $B\to D_s \pi$ decay from CLEO, our method sets an 
upper limit  $\left| \vub/\vcb \right| < 0.13$ which is very close 
to  the measured values from the semileptonic $B$ decays. 
We also discuss the possible breaking of factorization assumption.
 
\end{abstract} 
\end{titlepage}  
 
\newpage 
 
\renewcommand{\thefootnote}{\alph{footnote}} 
 
\section{Introduction} 
 
\noindent 
A precise determination of the Cabibbo-Kobayashi-Maskawa (CKM) matrix 
elements  \cite{CKM} is one of the key issues in the study of $B$ mesons 
and $B$-factory experiments \cite{TDRs}.   
For instance, the value of $\vub$ 
being non-zero is a necessary condition for CP-violation to occur in 
the SM; otherwise, we have to seek for new physics explanation 
for the observed CP violation in the $K_L\to \pi \pi$ 
decays \cite{refCPfirst}.  For a stringent test of the SM and a search 
for new physics, it is important to make a precise measurement of the 
modulus $|\vub|$ from the decays of $B$ mesons. 
 
Theoretical and experimental studies on $\vub$ have been mostly 
focused on the semi-leptonic $B$ meson decays.  Observations of 
inclusive and exclusive semileptonic $b\to u$ transitions by the 
CLEO \cite{CLEO,BSemilforVub} and ARGUS \cite{ARGUS} 
experiments confirm that $\vub$ is indeed nonzero, but these 
measurements suffered from large uncertainty due to model-dependence in 
extracting the value of $\vub$.  For instance, determination of $\vub$ 
from the inclusive $b \to u \ell \nu$ using the leptonic end-point momentum 
involves large uncertainty in determining the fraction of 
partial decay width for $E_\ell$ being larger than some cut value. 
In particular, the dependences of the lepton energy spectrum on 
perturbative and non-perturbative QCD corrections as well as on the 
unavoidable model-specific parameters are the strongest at the end-point 
region, which makes the determination of 
$|\vub/\vcb|$ very difficult by this method.  There have been other 
suggestions for avoiding these difficulties in studying the inclusive 
semileptonic $b\to u$ decays, for example,  by using the 
invariant mass of the hadronic system  \cite{MXforVub,invmass} 
recoiling against $\ell \nu$, or the invariant lepton mass \cite{invlep}.  
Experimental studies have been made using the invariant mass of the hadronic 
system to separate $b\to u$ from $b \to c$ decays, and measure 
$|\vub/\vcb|$ \cite{LEPMXU}. 
By considering the exclusive semileptonic decay modes such as  
$B \to \pi \ell \nu$ and $\rho \ell \nu$, we may reduce the dependence on 
the $E_\ell$ spectrum, but we are confronted with different aspects of 
theoretical uncertainties.  In this case, determination of $\vub$ 
becomes very sensitive to the fraction of such exclusive decays with 
respect to the inclusive $b \to u \ell \nu$ and it has large 
uncertainty coming from hadronic form factors. 
 
Although traditional difficulties with the understanding of 
non-leptonic weak decays have prevented their use in determination of 
CKM elements, the possibility of measuring $|\vub|$ via non-leptonic 
decays of $B$ mesons to exclusive two meson final states 
\cite{koide,nonlepton} has been theoretically explored.  To avoid the 
theoretical difficulties of non-spectator decay diagrams, only those 
final states have to be chosen in which no quark and anti-quark  
($q \bar q$) pair has the same flavor \cite{nonlepton}.   
 
Within the factorization approximation\footnote{
For possible breaking of the factorization approach, we give a short
discussion in Section II.}
and after considering the final state interactions, exclusive two body decay 
modes of $B$ mesons would certainly be worthy of full investigation. 
In Ref.~ \cite{koide}, it was pointed out that for the extraction of 
$|\vub/\vcb|$ from nonleptonic $B$ meson decay data, study of the 
ratio ${\cal B}(B^0\to D_s^+\pi^-)/{\cal B}(B^0\to D_s^+ D^-)$ is useful.   
Since 
the final states $\pi^-D_s^+$ ($D^-D_s^+$) consist of a single isospin 
component $I=1$ ($I=1/2$), the decay amplitudes are independent of the 
phase shift caused by elastic rescatterings in final 
states.  Furthermore, the decay mode $B^0\to D_s^+\pi^-$ is caused by 
only one diagram, the $b\to u$ tree transition with external 
$W$ emission and so completely independent of the penguin-type 
interaction.  On the other hand, in the decay mode $B^0\to D_s^+ D^-$, 
although the dominant contribution is from the $b\to c$ tree 
diagram, $b\to s$ penguin diagrams also contribute.  In 
Ref.~ \cite{koide} the authors neglected the penguin effect by expecting its 
size to be small compared to that of the tree diagram.   
However, we find that the theoretical estimate of the penguin 
correction is not small enough  to be simply neglected for  
$B^0\to D_s^+ D^-$ decay mode.   
On the other hand, we find that  
$B^0\to D_s^+ D^{*-}$ decay mode receives much smaller penguin corrections 
compared to $B^0\to D_s^+ D^-$ decay mode. 
In this work we investigate more thoroughly the possibility 
of extracting $|\vub/\vcb|$ from study of the nonleptonic decay ratio  
${\cal B}(\overline{B}^0\to D_s^{(*)-}(\pi^+,\rho^+))/ 
{\cal B}(\overline{B}^0\to D_s^{(*)-}D^{(*)+})$. 
 
\newpage 
 
\section{Theory on $\overline{B}^0\to D_s^{(*)-}(\pi^+,\rho^+)$  
and $\overline{B}^0\to D_s^{(*)-}D^{(*)+}$ \\ 
and numerical analysis} 
 
\noindent 
While only tree diagrams contribute to $\overline{B}^0\to D_s^-\pi^+$ decay 
in the SM, the decays $\overline{B}^0\to D_s^-D^{(*)+}$ receive penguin 
contributions as well.  The relevant $\Delta B=1$ effective 
Hamiltonian has the form; 
\be 
{\cal H}^q_{\rm eff}=\frac{G_F}{\sqrt{2}} 
\left\{ V_{qb}V^*_{cs}\Big[ c_1(\mu)O^q_1(\mu)+c_2(\mu)O^q_2(\mu) \Big] 
       -V_{tb}V^*_{ts}\sum^{10}_{i=3}c_i(\mu)O_i(\mu) \right\} 
+ {\rm h.c.}, 
\ee 
where $O_{1,2}$ represent QCD corrected tree-level operators and  
$O_{3-6}$ ($O_{7-10}$) 
the QCD (electroweak) penguin operators, which are defined as 
\be 
&&O^q_1=\bar{q}\g_\mu L b \bar{s}\g^\mu L c, \qquad\qquad 
  O^q_2=\bar{q}_\alpha\g_\mu L b_\beta \bar{s}_\beta\g^\mu L c_\alpha, 
\nonumber\\ 
&&O_{3(5)}=\bar{s}\g_\mu L b  
          \bar{c}\g^\mu L(R) c, \qquad   
  O_{4(6)}=\bar{s}_\alpha\g_\mu L b_\beta  
          \bar{c}_\beta\g^\mu L(R) c_\alpha,\\ 
&&O_{7(9)}=\bar{s}\g_\mu L b  
       \bar{c}\g^\mu R(L) c, \qquad   
  O_{8(10)}=\bar{s}_\alpha\g_\mu L b_\beta  
         \bar{c}_\beta\g^\mu R(L) c_\alpha,\nonumber 
\ee 
with $R(L)\equiv 1\pm\g_5$ and $q=u$ ($c$) for  
${\overline B}\to D_s \pi$ (${\overline B}\to D_s D^{(*)}$) 
decays. 
 
In the factorization approximation, the decay amplitudes of our 
interest are expressed as 
\be 
A(\overline{B}^0\to D_s^{(*)-}(\pi^+,\rho^+))= 
\frac{G_F}{\sqrt{2}}V_{ub}V_{cs}^*a_1 
            \langle D_s^-|\bar{s}\g^\mu L c|0\rangle  
         \langle \pi^+,\rho^+|\bar{u}\g_\mu L b|\overline{B}^0\rangle , 
\ee 
\be 
A(\overline{B}^0\to D_s^-D^+)&=&\frac{G_F}{\sqrt{2}}\left\{ 
      V_{cb}V_{cs}^*a_1-V_{tb}V_{ts}^*\Big[a_4+a_{10} 
   +2(a_6+a_8)\frac{m^2_{D_s}}{(m_b-m_c)(m_c+m_s)}\Big] \right\}\nonumber\\ 
         &&   \times\langle D_s^-|\bar{s}\g^\mu L c|0\rangle  
           \langle D^+|\bar{c}\g_\mu L b|\overline{B}^0\rangle , 
\ee 
\be 
A(\overline{B}^0\to D_s^-D^{*+})&=&\frac{G_F}{\sqrt{2}}\left\{ 
         V_{cb}V_{cs}^*a_1-V_{tb}V_{ts}^*\Big[a_4+a_{10} 
     -2(a_6+a_8)\frac{m^2_{D_s}}{(m_b+m_c)(m_c+m_s)}\Big] \right\}\nonumber\\ 
             &&   \times\langle D_s^-|\bar{s}\g^\mu L c|0\rangle  
                  \langle D^{*+}|\bar{c}\g_\mu L b|\overline{B}^0\rangle , 
\ee 
\be 
A(\overline{B}^0\to D_s^{*-}D^+)&=&\frac{G_F}{\sqrt{2}} 
    \left\{ 
         V_{cb}V_{cs}^*a_1-V_{tb}V_{ts}^*(a_4+a_{10})\right\} 
         \langle D_s^{*-}|\bar{s}\g^\mu L c|0\rangle  
         \langle D^+|\bar{c}\g_\mu L b|\overline{B}^0\rangle  
\ee 
Here $a_j$'s represent effective parameters defined as 
\be 
a_{2i}=c_{2i}^{\rm eff}+\frac{1} 
{(N_c^{\rm eff})_{2i}}c_{2i-1}^{\rm eff},\quad 
a_{2i-1}=c_{2i-1}^{\rm eff}+\frac{1}{(N_c^{\rm eff})_{2i-1}}c_{2i}^{\rm eff}, 
\label{Nc} 
\ee 
where $c_i^{\rm eff}$ are the renormalization scheme and scale 
independent effective Wilson coefficients, and $N_c^{\rm eff}$ is the 
so-called effective color number, which is supposed to include 
non-factorization effects as well as color suppression effect, and  
can thus be considered a free parameter. 
 
Using $V_{tb}V_{ts}^*\cong -V_{cb}V_{cs}^*$, 
one can cast the amplitudes in more compact forms 
\be 
A(\overline{B}^0\to D_s^{(*)-}D^{(*)+})&=& 
      \frac{G_F}{\sqrt{2}} 
         V_{cb}V_{cs}^*\tilde{a}_1(B\to D^{(*)}D_s^{(*)}) 
          \langle D_s^{(*)-}|\bar{s}\g^\mu L c|0\rangle  
                  \langle D^{(*)+}|\bar{c}\g_\mu L b|\overline{B}^0\rangle , 
\ee 
where 
\be 
\tilde{a}_1(B\to DD_s)&=&a_1\left(  
          1+\frac{a_4+a_{10}}{a_1} 
 +2\frac{a_6+a_8}{a_1}\frac{m^2_{D_s}}{(m_b-m_c)(m_c+m_s)}\right),\nonumber\\ 
\tilde{a}_1(B\to D^*D_s)&=&a_1\left(  
            1+\frac{a_4+a_{10}}{a_1} 
 -2\frac{a_6+a_8}{a_1}\frac{m^2_{D_s}}{(m_b+m_c)(m_c+m_s)}\right),\nonumber\\ 
\tilde{a}_1(B\to DD_s^*)&=&a_1\left(  
            1+\frac{a_4+a_{10}}{a_1}\right). 
\ee 
Using the numerical values of $a_j$'s in Ref.~ \cite{cheng}, 
the effective parameters $\tilde{a}_1$ defined above are related to $a_1$ by 
\be 
|\tilde{a}_1(B\to DD_s)|&=&0.847a_1,\nonumber\\ 
|\tilde{a}_1(B\to D^*D_s)|&=&1.037a_1,\nonumber\\ 
|\tilde{a}_1(B\to DD_s^*)|&=&0.962a_1. 
\ee 
From the above relations one can see that, at the amplitude level, 
the penguin contributions to 
$\overline{B}^0\to D_s^-D^{*+}$ decay ($3.7\%$) are much smaller than those for 
$\overline{B}^0\to D_s^-D^+$ mode ($15.3\%$).  Actually the penguin effects on 
$\overline{B}^0\to D_s^-D^+$ decay are not small enough to be simply 
neglected.  As mentioned in the Introduction, the penguin effects 
are neglected in Ref.~ \cite{koide}.  We note that for 
$\overline{B}^0\to D_s^-D^{*+}$ decay mode the penguin contribution can 
be neglected. 
This difference of penguin contributions to the similar modes 
$\overline{B}^0\to D_s^-D^+$ and $\overline{B}^0\to D_s^-D^{*+}$ 
is due to the different chiral structure of the final states. 
$B\to D^*$ transitions occur through axial vector currents, 
while $B\to D$ through vector currents.  
 
Then, the ratios 
\be 
{\cal R}_{(\pi,\rho)/D^{(*)}}&\equiv&  
    \frac{{\cal B}(\overline{B}^0\to D_s^-(\pi^+,\rho^+))} 
{{\cal B}(\overline{B}^0\to D_s^-D^{(*)+})}\\ 
{\rm and}~~~~\tilde{\cal R}_{\pi/D}&\equiv&  
    \frac{{\cal B}(\overline{B}^0\to D_s^{*-}\pi^+)}{{\cal B}(\overline{B}^0\to D_s^{*-}D^+)} 
\ee 
are given as 
\be 
{\cal R}_{\pi/D}=\left|\frac{\vub}{\vcb}\right|^2 
\left(\frac{a_1}{\tilde{a}_1(B\to DD_s)}\right)^2 
           \frac{(m_B^2-m_\pi^2)^2}{(m_B^2-m_D^2)^2} 
           \left(\frac{p_c^\pi}{p_c^D}\right) 
           \left(\frac{F_0^{B\pi}(m_{D_s}^2)}{F_0^{BD}(m_{D_s}^2)}\right)^2, 
\label{RpiD} 
\ee 
\be 
{\cal R}_{\rho/D}=\left|\frac{\vub}{\vcb}\right|^2 
\left(\frac{a_1}{\tilde{a}_1(B\to DD_s)}\right)^2 
           \frac{m_B^2}{(m_B^2-m_D^2)^2} 
           \left(\frac{p_c^{\rho 3}}{p_c^D}\right) 
         \left(\frac{2A_0^{B\rho}(m_{D_s}^2)}{F_0^{BD}(m_{D_s}^2)}\right)^2, 
\label{RrhoD} 
\ee 
\be 
{\cal R}_{\pi/D^*}=\left|\frac{\vub}{\vcb}\right|^2 
\left(\frac{a_1}{\tilde{a}_1(B\to D^*D_s)}\right)^2 
           \frac{(m_B^2-m_\pi^2)^2}{m_B^2} 
           \left(\frac{p_c^\pi}{p_c^{D^*3}}\right) 
       \left(\frac{F_0^{B\pi}(m_{D_s}^2)}{2A_0^{BD^*}(m_{D_s}^2)}\right)^2, 
\label{RpiDST} 
\ee 
\be 
{\cal R}_{\rho/D^*}=\left|\frac{\vub}{\vcb}\right|^2 
\left(\frac{a_1}{\tilde{a}_1(B\to D^*D_s)}\right)^2 
           \left(\frac{p_c^\rho}{p_c^{D^*}}\right)^3 
        \left(\frac{A_0^{B\rho}(m_{D_s}^2)}{A_0^{BD^*}(m_{D_s}^2)}\right)^2, 
\label{RrhoDST} 
\ee 
\be 
\tilde{\cal R}_{\pi/D}=\left|\frac{\vub}{\vcb}\right|^2 
\left(\frac{a_1}{\tilde{a}_1(B\to DD_s^*)}\right)^2 
           \left(\frac{p_c^\pi}{p_c^D}\right)^3 
       \left(\frac{F_1^{B\pi}(m_{D_s^*}^2)}{F_1^{BD}(m_{D_s^*}^2)}\right)^2, 
\label{RD} 
\ee 
where $p_c^X$ is the c.m. momentum of the decay particle $X$. 
Here the form factors follow the following parameterization   \cite{BSW}: 
\be 
\langle P^\prime (p^\prime) |V_\mu| P(p)\rangle &=&  
    \left(p_\mu+p^\prime_\mu-\frac{m_P^2-m_{P^\prime}^2}{q^2}\right)F_1(q^2) 
             +\frac{m_P^2-m_{P^\prime}^2}{q^2}q_\mu F_0(q^2),\nonumber\\ 
\langle V(p^\prime,\epsilon) |V_\mu| P(p)\rangle &=& 
       \frac{2}{m_P+m_V}\epsilon_{\mu\nu\alpha\beta} 
           \epsilon^{*\nu}p^\alpha p^{\prime\beta}V(q^2),\nonumber\\ 
\langle V(p^\prime,\epsilon) |A_\mu| P(p)\rangle &=& 
       i\left[(m_P+m_V)\epsilon_\mu A_1(q^2)-\frac{\epsilon\cdot p}{m_P+m_V} 
              (p+p^\prime)_\mu A_2(q^2)\right.\nonumber\\ 
    &&\left.-2m_V\frac{\epsilon\cdot p}{q^2}q_\mu [A_3(q^2)-A_0(q^2)]\right], 
\ee 
where $q=p-p^\prime$, $F_1(0)=F_0(0)$, $A_3(0)=A_0(0)$, 
$$A_3(q^2)=\frac{m_P+m_V}{2m_V}A_1(q^2)-\frac{m_P-m_V}{2m_V}A_2(q^2),$$ 
and $P$, $V$ denote the pseudoscalar and vector mesons, respectively. 
 
As can be inferred from the explicit expressions in Eqs.~(\ref{RpiD}) -- 
(\ref{RD}), these ratios do not have much dependence on values of 
$a_i$.  Especially, for $\overline{B}^0\to D_s^-D^{*+}$ decay, the penguin 
contributions add up destructively and so the 
dependence on $a_i$ is almost cancelled out in the ratio.  However, explicit 
calculations for the ratios of branching fractions depend strongly on 
the form factors.  In the following analysis,  
we consider five models for the form factors of $B\to \pi$  
transitions: two quark-model 
approaches (the Bauer-Stech-Wirbel (BSW) model   \cite{BSW} and 
Melikhov/Beyer  \cite{MB}), light-cone sum rules (LCSR \cite{LCSR}), 
lattice QCD (UKQCD \cite{UKQCD}), and relativistic light-front (LF) 
quark model  \cite{LF}.  And for $B\to D^{(*)}$ transitions, we adopt 
BSW, Melikhov/Stech  \cite{MS}, and LF models.  In Table 1, we list 
explicit numerical values of form factors evaluated at 
$q^2=m_{D_s^{(*)}}^2$.  Then we get the theoretical predictions 
\be 
{\cal R}_{\pi/D} \equiv   
    \frac{{\cal B}(\overline{B}^0\to D_s^-\pi^+)}{{\cal B}(\overline{B}^0\to D_s^-D^{+})} 
           &=& 0.424\left|\frac{\vub}{\vcb}\right|^2 
               \left[\frac{F_0^{B\pi}(m_{D_s}^2)}{0.319}\right]^2 
               \left[\frac{0.740}{F_0^{BD}(m_{D_s}^2)}\right]^2\nonumber\\ 
           &=& [0.424\pm 0.041]\left|\frac{\vub}{\vcb}\right|^2,\label{Eq1} 
\ee 
\be 
{\cal R}_{\rho/D} \equiv   
    \frac{{\cal B}(\overline{B}^0\to D_s^-\rho^+)}{{\cal B}(\overline{B}^0\to D_s^-D^{+})} 
           &=& 0.443\left|\frac{\vub}{\vcb}\right|^2 
               \left[\frac{A_0^{B\rho}(m_{D_s}^2)}{0.398}\right]^2 
               \left[\frac{0.740}{F_0^{BD}(m_{D_s}^2)}\right]^2\nonumber\\ 
           &=& [0.443\pm 0.063]\left|\frac{\vub}{\vcb}\right|^2, 
\ee 
\be 
{\cal R}_{\pi/D^*} \equiv   
    \frac{{\cal B}(\overline{B}^0\to D_s^-\pi^+)}{{\cal B}(\overline{B}^0\to D_s^-D^{*+})} 
           &=& 0.459\left|\frac{\vub}{\vcb}\right|^2 
               \left[\frac{F_0^{B\pi}(m_{D_s}^2)}{0.319}\right]^2 
               \left[\frac{0.793}{A_0^{BD^*}(m_{D_s}^2)}\right]^2\nonumber\\ 
           &=& [0.459\pm 0.076]\left|\frac{\vub}{\vcb}\right|^2, 
\ee 
\be 
{\cal R}_{\rho/D^*} \equiv   
    \frac{{\cal B}(\overline{B}^0\to D_s^-\rho^+)}{{\cal B}(\overline{B}^0\to D_s^-D^{*+})} 
           &=& 0.480\left|\frac{\vub}{\vcb}\right|^2 
               \left[\frac{A_0^{B\rho}(m_{D_s}^2)}{0.398}\right]^2 
               \left[\frac{0.793}{A_0^{BD^*}(m_{D_s}^2)}\right]^2\nonumber\\ 
           &=& [0.480\pm 0.094]\left|\frac{\vub}{\vcb}\right|^2, 
\ee 
\be 
\tilde{\cal R}_{\pi/D} \equiv   
    \frac{{\cal B}(\overline{B}^0\to D_s^{*-}\pi^+)}{{\cal B}(\overline{B}^0\to D_s^{*-}D^{+})} 
           &=& 0.456\left|\frac{\vub}{\vcb}\right|^2 
               \left[\frac{F_1^{B\pi}(m_{D_s^*}^2)}{0.367}\right]^2 
               \left[\frac{0.817}{F_1^{BD}(m_{D_s^*}^2)}\right]^2\nonumber\\ 
           &=& [0.456\pm 0.038]\left|\frac{\vub}{\vcb}\right|^2,\label{Eq2} 
\ee 
where the errors are originated only from the dependence on the  
hadronic form-factors. 
 
Considering the current experimental results \cite{PDG,APS2K} 
\be 
&&{\cal B}(\overline{B}^0\to D_s^-\pi^+)<5.1\times 10^{-5},\nonumber\\ 
&&{\cal B}(\overline{B}^0\to D_s^-\rho^+)<7.0\times 10^{-4},\nonumber\\ 
&&{\cal B}(\overline{B}^0\to D_s^-D^+)=(8.0\pm 3.0)\times 10^{-3},\nonumber\\ 
&&{\cal B}(\overline{B}^0\to D_s^-D^{*+})=(9.6\pm 3.4)\times 10^{-3},\nonumber\\ 
&&{\cal B}(\overline{B}^0\to D_s^{*-}D^+)=(1.0\pm 0.5)\times 10^{-2},\nonumber\\ 
&&{\cal B}(\overline{B}^0\to D_s^{*-}\pi^+)<7.5\times 10^{-5}, 
\label{bound} 
\ee 
we then estimate 
\be 
\left|\frac{\vub}{\vcb}\right| < \left\{  
\begin{array}{l} 
  0.13 \quad{\rm from}\quad {\cal R}_{\pi/D}, \\ 
  0.12 \quad{\rm from}\quad {\cal R}_{\pi/D^*}, \\ 
  0.14 \quad{\rm from}\quad \tilde{\cal R}_{\pi/D},\\ 
  0.48 \quad{\rm from}\quad {\cal R}_{\rho/D}, \\ 
  0.44 \quad{\rm from}\quad {\cal R}_{\rho/D^*}. 
\end{array}\right.  
\ee 
We note that 
with the new preliminary upper limits on $B\to D_s \pi^+$ from 
CLEO \cite{APS2K}, the upper limit on $\left|{\vub}/{\vcb}\right|$ 
is already very close to the current estimate 
$\left|{\vub}/{\vcb}\right| = 0.09\pm 0.025$ \cite{Falk99}. 
 
Next we consider direct extraction of $|\vub|$ from the  
$\overline{B}^0\to D_s^- \pi^+$ decay rate. 
As mentioned earlier, this mode is one of the cleanest processes 
without any penguin corrections. 
Main uncertainties come from the form factor $F^{B\pi}_0$ and 
Wilson coefficients or the effective parameter $a_1$. 
The decay rate is given as 
\be 
\Gamma(\overline{B}^0 &\to& D_s^- \pi^+) = \frac{G_F^2}{2}|\vub\vcs^*|^2 
     \frac{p^\pi_c}{8\pi m_B^2}(m_B^2-m_\pi^2)^2[a_1 f_{D_s} 
      F^{B\pi}_0(m_{D_s}^2)]^2\nonumber\\ 
    &=& (1.065\times 10^{-12})|\vub|^2 
    \left[\frac{a_1}{1.059}\right]^2 
  \left[\frac{f_{D_s}}{0.240\;{\rm GeV}}\right]^2
    \left[\frac{F^{B\pi}_0(m_{D_s}^2)}{0.319}\right]^2\;{\rm GeV} .
\ee 
%
In order to estimate theoretical uncertainty from the effective coefficient 
$a_1$, we choose different values for $N_c^{\rm eff}$, the effective number 
of colors appearing in Eq.~(\ref{Nc}), $N_c^{\rm eff}=$ 2, 3 and $\infty$, 
and we use for the renormalization scheme independent Wilson coefficients 
$c_1^{\rm eff}=1.149$ and $c_2^{\rm eff}=-0.325$  \cite{cheng}. 
Considering the new preliminary upper limits on 
$\overline{B}^0 \to D_s^- \pi^+$ 
in Eq.~(\ref{bound}), we then estimate 
\be 
|\vub| < (0.450 \pm 0.020 \pm 0.021)\times 10^{-2} , 
\ee 
where the first error is due to the effective coefficient $a_1$ 
and the second one is from the dependence on the form factor $F^{B\pi}_0$. 

Finally we note on the possible breaking \cite{Beneke}
of the factorization assumption, on which our previous results are based;
we now generalize Eqs. (\ref{RpiD}) -- (\ref{RD}) by including
the factor for breaking of factorization,
\be
{\cal R}_{i/J} \equiv   
  \frac{{\cal B}(\overline{B}^0\to D_s^{(*)-}i^+)}
       {{\cal B}(\overline{B}^0\to D_s^{(*)-}J^{+})}  =
\left|\frac{\vub}{\vcb}\right|^2  K_{i/J} \left[\frac{a_1}{\tilde{a}_1}\right]^2_{i/J} 
    \left[\frac{F^{B \to i}(m_{D_s^{(*)}}^2)}
           {F^{B \to J}(m_{D_s^{(*)}}^2)}\right]^2    
\left[ 1 + {\rm (F-B)}_{i/J} \right], \label{Eqx}  
\ee 
where $i=\pi,\rho$ and $J=D,D^*$. $K_{i/J}$ is the kinematic
phase space factor, and $a_1/\tilde{a}_1$ is the ratio of effective Wilson
coefficients, for which we used values estimated within factorization
assumption.  Because the decay $\overline{B}^0\to D_s^{(*)-}i^+$ is only
through tree diagrams and $\overline{B}^0\to D_s^{(*)-}J^{+}$ is polluted
by penguin diagrams, the ratios ${\cal R}_{i/J}$ would possibly be the best
observables to measure the breaking of factorization,
(F--B)$_{i/J}$.  Instead of measuring $|V_{ub}/V_{cb}|$ as proposed, if we
use the value of $|V_{ub}/V_{cb}|$ measured from the semileptonic decays of
$B$ mesons,  then we can systematically estimate the (F--B)$_{i/J}$
through experimental values of ${\cal R}_{i/J}$.  

\section{Discussions on experimental feasibility and Summary}  
 
\noindent 
We have investigated the possibility of extracting $|\vub/\vcb|$ and  
$|\vub|$ from non-leptonic exclusive decays of $B$ meson  
into two meson final states.   
In particular, we calculated the ratios ${\cal B}(\overline{B}^0\to 
D_s^{(*)-}(\pi^+,\rho^+))/{\cal B}(\overline{B}^0\to D_s^{(*)-}D^{(*)+})$ 
including penguin effects in the factorization assumption.  By taking 
the ratios, some model-dependence on the coefficients $a_i$ and 
hadronic form-factors is reduced.  We found that the $\overline{B}^0\to
D_s^-D^{+}$  mode has considerable penguin corrections ($\sim 15\%$ at the
amplitude  level) which cannot be simply ignored as it was done in  Ref.~ 
\cite{koide}.  We also found that the $\overline{B}^0\to D_s^-D^{*+}$ mode has
very small penguin  corrections.  On the other hand, the $\overline{B}^0\to
D_s^-D^{+}$ mode  has smaller form-factor dependence than $\overline{B}^0\to
D_s^-D^{*+}$.  Therefore, these modes can complement each other in measuring 
$\vub/\vcb$, in conjunction with $B\to D_s^{(*)} \pi$ and  
$B \to D_s \rho$. 
 
We have shown that the relevant hadronic form factor uncertainties in our 
methods are typically at the level of 15\% (Eqs.~(\ref{Eq1}) -- 
(\ref{Eq2})).   On the other hand, the model-dependence uncertainties in 
the measurement of $\left|\vub\right|$ from exclusive semileptonic $B$ 
decays are currently at the level of 20\%  \cite{BSemilforVub}. 
Therefore, if we can contain the experimental uncertainty of our 
method within 15\%, the method described in this paper becomes 
competitive with the semileptonic analyses.  This implies that we have 
to determine the decay branching ratios of $B\to D_s^{(*)} \pi$, 
{\it etc.} with experimental uncertainties being less than 30\%. 
 
Current experimental uncertainties for the branching ratios of  
$B \to D_s D^{*}$ decays are at the level of approximately 30\%. 
The recent CLEO measurements \cite{CLEOprd53p4734} of 
${\cal B}(\overline{B}^0\to D_s^- D^+) = (8.4 \pm 3.0) \times 10^{-3}$ and 
${\cal B}(\overline{B}^0\to D_s^- D^{*+}) = (9.0 \pm 2.7) \times 10^{-3}$ are 
based on the event sample of $(2.19\pm 0.04)\times 10^6$ $\BBbar$ 
pairs.  With high-statistics event sample expected from the 
$B$-factories, we anticipate reducing the uncertainties of these decays 
to below 10\% level in the very near future. 
As for the decay $B \to D_s^{(*)} \pi$ and $D_s \rho$, none of them 
have been experimentally measured yet.  Recently, CLEO has presented 
the following preliminary upper limits: 
${\cal B}(B \to D_s \pi^+) < 8.9 \times 10^{-5}$ and  
${\cal B}(B \to D_s^{*} \pi^+) < 7.5 \times 10^{-5}$   \cite{APS2K},  
based on an event sample of $9.7\times 10^6$ 
$\BBbar$ pairs.  We expect to have larger amount of data from each 
$B$-factory experiment very soon.  This, combined with much improved 
hadron identification and vertexing capabilities of the $B$-factory 
experiments, will improve the sensitivity of $B \to D_s^{(*)} \pi$ and 
$D_s \rho $ searches down to ${\cal B} \approx 10^{-5}$ level and discovery 
of the $B \to D_s^{(*)} \pi (\rho)$ reactions might be possible in the 
near future. 
 
Using the current estimate of  
$\left|{\vub}/{\vcb}\right|=0.09\pm 0.025$  \cite{Falk99},  
along with the world-average value of  
${\cal B}(\overline{B}^0\to D_s^-D^+)=(8.0\pm 3.0)\times 10^{-3}$   \cite{PDG},  
we take ${\cal B}(\overline{B}^0\to D_s^-\pi^+) = 2.7\times 10^{-5}$ 
as the expected branching ratio.  Based on this, we estimate experimental 
conditions to achieve 30\% statistical uncertainty in  
${\cal B}(\overline{B}^0\to D_s^-\pi^+)$ from the $B$-factories.  Consider we 
measure $B\to D_s \pi$ decay through $D_s \to \phi \pi$ and  
$\phi \to K^+ K^-$.   
In terms of signal detection efficiency $\varepsilon$ and 
integrated luminosity $\cal{L}$ (in fb$^{-1}$), the expected number of 
reconstructed $B\to D_s \pi$ events are 
\be  
N &=& \varepsilon ~{\cal L} \times 10^6 \times 2.7\times 10^{-5}  
 \times 0.018\nonumber\\ 
  &=& 0.5 ~\varepsilon ~{\cal L} ,\nonumber 
\ee 
where 0.018 is the decay branching ratio of the particular $D_s$ decay 
mode that we consider.  Assuming $\varepsilon=17\%$   \cite{BELLEDsPi} we 
need ${\cal L}\approx 120~{\rm fb}^{-1}$ to obtain 10 signal events. 
We can further improve the experimental sensitivity by several 
factors, if we include other decay channels of $D_s$ and if we also 
analyze related modes such as  
$B\to D_s^* \pi$, $B\to D_s^{(*)} \rho$, $B\to D_s^{(*)} \omega$, {\it etc.}   
In this case, with ${\cal L}\approx 50~{\rm fb}^{-1}$ of data from the 
$B$-factories, it may be possible to obtain a competitive, independent 
measurement of $\vub$ from exclusive non-leptonic $B$ decays. 
 
There have been other studies of using nonleptonic $B$ decays for 
measuring $|V_{ub}/V_{cb}|$.  For example, there were studies of 
utilizing fully inclusive nonleptonic $b\to u\bar{c}s'$ 
decays~ \cite{fullincb2u} or exploiting semi-inclusive nonleptonic $B$ 
decays $B\to D_s^+ X_u$~ \cite{semiincb2dsxu}.  The method suggested in 
Ref.~ \cite{fullincb2u} is clean in theoretical calculation.  But it is 
based on the assumption that we can experimentally separate, without 
introducing much model-dependent uncertainty, $b\to u\bar{c}s'$ 
decays from the dominant $b\to c\bar{c}s'$ decays which is 
two-orders-of-magnitude larger in size.  With existing experimental 
capabilities of $B$-factory experiments, this assumption is not 
justified.  Ref.~ \cite{semiincb2dsxu}  considers more carefully 
the experimental backgrounds, but ignores the 
significance of the continuum background which can be statistically dominant 
in the interesting range of $2.0<p_{D_s}<2.5$~GeV/c   \cite{Menary}.

\section*{Acknowledgments} 
 
\noindent 
We thank M. Beneke and G. Cvetic for careful reading of the manuscript and  
their valuable comments. 
The work of C.S.K. was supported  
in part by  Grant No. 2000-1-11100-003-1 and SRC Program of the KOSEF, 
and in part by the KRF Grant, Project No. 2000-015-DP0077. 
The work of Y.K. was supported by Brain Korea 21 Project. 
The work of J.L. was supported in part by the KRF Grant, Project No. 1997-011-D00015, 
and in part by Research Intern Program of the KOSEF. 
The work of W.N. was supported by the BSRI Program of MOE, Project No. 
99-015-DI0032. 
 
\begin{table} 
{Table~1}. {Numerical values of form factors at  
 $q^2=m_{D_s^{(*)}}^2$ in various  form-factor models}\par 
\begin{center} 
\begin{tabular}{|c|c|c|c|c|c|c|} 
{} & $F_0^{B\pi}(m_{D_s}^2)$ & $F_1^{B\pi}(m_{D_s^*}^2)$ &  
$F_0^{BD}(m_{D_s}^2)$  
   & $F_1^{BD}(m_{D_s^*}^2)$ & $A_0^{B\rho}(m_{D_s}^2)$&  
$A_0^{BD^*}(m_{D_s}^2)$\\ 
\hline 
BSW & 0.377 & 0.395 & 0.753 & 0.776 & 0.326 & 0.690\\ 
Beyer/Melikhov & 0.299 & 0.357 & $\cdot$ & $\cdot$ & 0.381 & $\cdot$\\ 
LCSR & 0.322 & 0.382 & $\cdot$ & $\cdot$ & 0.457 & $\cdot$\\ 
UKQCD & 0.301 & 0.359 & $\cdot$ & $\cdot$ & 0.469 & $\cdot$\\ 
Melikhov/Stech & $\cdot$ & $\cdot$ & 0.722 & 0.806 & $\cdot$ & 0.818\\ 
LF & 0.295 & 0.349 & 0.746 & 0.869 & 0.357 & 0.872\\ 
\end{tabular} 
\end{center} 
\end{table} 

\end{document}